\documentclass[sigconf]{acmart}
\AtBeginDocument{%
  }

\setcopyright{acmlicensed}
\copyrightyear{2025}
\acmYear{2025}
\begin{document}

\title{Exploring Multiview UI Layouts and Placement Strategies for Collaborative Sensemaking in Virtual Reality}


\author{Tamzid Hossain}
\authornotemark[1]
\affiliation{%
  \institution{Shahjalal University of Science and Technology}
  \city{Sylhet}
  \country{Bangladesh}
}
\email{tamzid.cse.sust@gmail.com}

\author{Md. Fahimul Islam}
\authornotemark[1]
\affiliation{%
  \institution{Shahjalal University of Science and Technology}
  \city{Sylhet}
  \country{Bangladesh}}
\email{fahimulislamfahim11701068@gmail.com}

\author{Farida Chowdhury}
\affiliation{%
  \institution{BRAC University}
  \city{Dhaka}
  \country{Bangladesh}}
\email{farida.chowdhury@bracu.ac.bd}


\renewcommand{\shortauthors}{Hossain et al.}

\begin{abstract}
   Immersive technologies expand the potential for collaborative sense-making and visual analysis via head-worn displays (HWDs), offering customizable, high-resolution perspectives of a shared visualization space. In such an immersive environment, window/view management is crucial for collaborative sense-making tasks. However, the role of document types (graphs, images) and pair dynamics in collaborative layout formation has rarely been explored. We conducted a user study with 20 participants to explore how pair of users organize multiview windows in remote immersive workspaces during tasks such as search, comparison, and classification. Findings show that users often arrange windows in a semi-circular layout for pair collaboration. Image+text documents reduce mental and temporal demand in comparison tasks, while graphs lower task load for classification. Conflicts in window selection arise mainly in complex comparisons, with frequent discussion and reorganization during difficult tasks. Based on these insights, we propose design guidelines for multiview systems that support VR collaboration and brainstorming.
\end{abstract}



\begin{teaserfigure}
  \includegraphics[width=\textwidth]{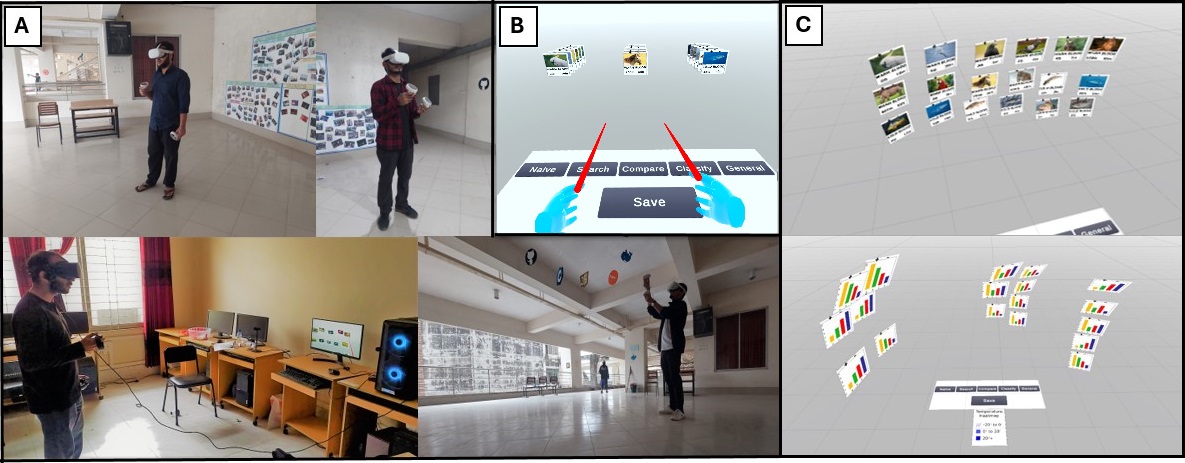}
  \caption{Collaboration and brainstorming sessions are crucial in modern VR derived world. We explore different aspect of interacting with multiview UI layouts in such scenarios.}
  \Description{Enjoying the baseball game from the third-base
  seats. Ichiro Suzuki preparing to bat.}
  \label{fig:teaser}
\end{teaserfigure}


\maketitle

\section{Introduction} 

Immersive technologies have opened up new possibilities for collaborative sense-making, enabling users to work together in immersive, three-dimensional environments. Collaborative brainstorming and sensemaking activities are nearly always required to solve real-world problems, such as carrying out scientific research, design, crime investigation, etc. These collaborative efforts can be enhanced by immersive technologies due to their ability to provide an expanded display area, facilitate collaboration, and seamlessly integrate into existing workflows. Previous studies\cite{SpaceToThink, SpatialLabeling, AnExaminationofGrouping} have demonstrated that users utilize space and location to arrange and systematize ideas and thoughts while generating sense. The convergence of these variables underscores the immense potential of utilizing virtual reality (VR) to facilitate collaborative ideation and sensemaking. Our motivation for the study is the growing interest in collaborative work using VR technologies, particularly in design, engineering, and education. However, while VR offers many advantages for collaboration, it also presents unique challenges, such as the need to design effective interfaces that can support multiple users and viewpoints.




Although multiview interfaces are useful for various purposes e.g. learning, data visualizarion and problem solving, there has been limited research on the organization of multiview content layouts in immersive space. Prior work explored multiview layouts for map exploration\cite{mapsAroundMe}, brainsorming\cite{10316471}, data transformation\cite{10197223} and multiple desktop application\cite{10536416} for single user in VR and XR environment. Influence of surrounding environment on spatial organization of virtual documents were also explored for AR\cite{WhereShouldWePutIt} and MR\cite{10.1145/3613905.3638181, 10.1145/3544549.3585853} context. But little is known about the multiview exploration for different types of visualization when pairs collaborative in a remote VR environment. 

In this paper, we explored how multiview documents are organized and placed for remote collaborative sense-making in VR. More specifically, we investigated the factors that should be taken into account while designing functional document layouts for collaborative tasks in VR. Given the freedom of layout configurations, we examined how pair of users arrange a series of windows in the immersive space. Our examination of layout geometry suggests that there are mainly two types of layout patterns (with variation in vertical and horizontal alignment) that groups prefer: planner and semi-circular. Of these two types of layouts, 55\% of people prefer the semi-circular layout (vertical + horizontal). Our usability study (NASA-TLX) reveal significant effect on mental and temporal demand and effort as task and document types change. Based on the task type, the results suggest that comparison type tasks are less challenging with image+text based documents, and graph visualization can be used for solving classification related problems. We also observed a significant number of conflicts as the complexity of the comparison task changed. Lastly, in our data analysis, we discovered several collaboration techniques within pairs. The majority of participant's time is spent in collaborative mode. The second most prevalent interaction technique is layout manipulation, which is driven by iterative collaborative brainstorming.


Through our work, we aim to produce the following new knowledge and contributions:
\begin{itemize}
    \item A study investigating multiview window layouts and management in VR for collaborative sensemaking task.
    \item In-depth understanding of window arrangement, positioning strategies, pair dynamics, and the associated influential factors.
    \item An analysis of task load index measurement for various document and task types, and complexities for remote sensemaking activities in VR.
    \item A comparative analysis of different collaboration strategies for pair in remote VR sensemaking.
    \item Guidelines for designing layouts of multiple views in an immersive (VR) collaborative environment based on document and problem types.

\end{itemize}

\section{Related Work}

Since our study is primarily concerned with multiview window layouts in VR, we begin by reviewing the literature on view and window management in immersive space. Subsequently, we discuss prior research on collaborative sense-making within immersive environments. 

\subsection{Multiview Management in Immersive Space}
As HWDs do not have physical display constraints, it means that designers can use the abundant space of the wearer's display to initiate 3D capabilities. Multiple virtual windows can float in a 3D immersive environment, much like a real-world multi-monitor workstation \cite{PartitioningDigitalWorlds}. So researchers proposed a system, Personal Cockpit, a unique and efficient design for mobile multitasking on head-worn displays \cite{personalcockpit}. Each individual window in the personal cockpit is called a small multiple. The user's comfortability and lower task cost compelled researchers to explore small multiple windows in an immersive environment. For example, researchers proposed a system for visualizing small multiples in grid on a flat layout \cite{BentoBox} in 3D immersive environment. Researchers also used small multiples on ground where main focus was presented directly in front on the 3D trajectories \cite{FiberClay}.

Liu et al.\cite{DandEofsmallmultiples} explored the adaptation of 2D small-multiples visualisation on flat screens to 3D immersive spaces, and investigated design space for layouts of small multiples for data visualization. Their proposed prototype investigates a variety of interaction strategies for altering layouts in 3D environment and engaging with different types of data visualizations. Visual transition between horizontal tabletop display and vertical virtual display were also explored in recent studies, where they proposed guidelines to design interactive environment combining tabletop display and surrounding virtual display via see-through HMDs\cite{VisualTransition}. 

Using a combination of embedded visualization(in AR) and situated visualization (on a tablet), Zhau et al.\cite{zhou2024lights} observed improved user satisfaction and minimized attention-switching overheads in this setup.  Researchers \cite{liu2022effects} also explored the effect of virtual display layouts on spatial memory in visual analytical task, and showed show greater accuracy and positive rating towards flat layouts than circular-wraparound layouts when recalling spatial patterns.
 
Several studies have explored multiview positions and layouts in context of immersive situated analytics. Wen et al.\cite{wen2022effects} examined the effects of view layout for multi-view representations on situated analytics, where they analyzed the design tradeoffs for achieving high situatedness and effective analytics simultaneously. Researchers also investigated potential placements for data visualization in the context of outdoor exploration\cite{ghaemi2023visualization}. Li et al.\cite{li2023immerview} proposed flexible layout for multi-view visualizations that considers the spatial arrangement of objects in relation to the user’s position, viewing angle, and the object of focus.

The impact of physical surrounds and surfaces on multiple virtual document placement were also explored in the recent studies\cite{10.1145/3411763.3451588}\cite{10.1145/3613905.3638181}. Here they investigated how virtual content is placed in different AR/MR-enhanced room settings. In their recent work, Ellenberg et al.\cite{ellenberg2023spatiality} characterized the situated layouts of virtual contents by the level of spatial and semantic coupling.  Besides physical environment, Luo\cite{luo2024exploring} also considered another factor, people present. Unlike MapsAroundme\cite{mapsAroundMe} where multiview layout exploration were only investigated for map-exploration, Davidson et al.\cite{10316471} explored document layouts for multiple analysis tasks in immersive space, and provided guidelines for document-based sensemaking. In  Daeijavad et al.\cite{daeijavad2022layouts} argued that along with 4 dimensions that was described earlier by \cite{DandEofsmallmultiples}, the design space for layout should have interaction techniques as another dimension. In their work they added three more dimensions: Height, Detail level and Interaction. The organization of small multiples around the users have previously explored by several studies\cite{personalcockpit}\cite{liu2023datadancing}.

\subsection{Collaboration in Immersive Environment}
A collaborative virtual environment is a shared immersive environment where remote or co-located users can join, share information among themselves and interact in the 3D space\cite{CVE}. According to a recent survey, collaboration is becoming a key aspect in immersive environment research \cite{ImmersiveSurvey}. 
Though immersive environments can facilitate users in different ways, researchers see the greatest potential of immersive environments in collaborative workflow \cite{immersiveVisualization}\cite{yang2022towards}.

To facilitate remote side-by-side collaboration in immersive space, Sidenmark\cite{sidenmark2024desk2desk} presented an optimized hybrid workspaces of two collaborators, called Desk2Desk. The system adaptively merges dissimilar physical workspaces and adjusts each user’s workspace in layout and number of shared monitors, enssuring spatial consistency. Researchers \cite{collaboVR} have also presented a flexible framework, CollaboVR, that allows for multi-user communication in a VR immersive environment that can take place between users who are co-located as well as those who are geographically separated. This system provides shared freehand sketching and 2D sketches to 3D environment modelling, which encourages users to express their creative potential. In a recent study, Bovo et al.\cite{10536416} proposed WindowMirror, a toolkit that enables users to engage with various desktop programs simultaneously in a real-time XR environment. Remote immersive colaboration system for analyzing geo-spatial data was also developed in recent study\cite{Mahmood}.

Zaman et al.\cite{nRoom} conducted user studies on their proposed system to analyze the possible interaction strategies in a collaborative, immersive VR environment. Using their user study system (which allows multiple remote users to place different furniture in a VR room environment), the researchers proposed a guideline for a collaborative spatially enabled virtual reality GUI. Researchers\cite{CollabAR} also explored co-located group collaboration using mobile AR system and investigated the effect of task complexity on group interaction.

Previous research has been conducted on the designing of small multiple layouts and users' approaches for interacting with small multiples in immersive environments. When more than one participant from different locations joins a shared immersive environment for the purpose of collaborative information processing, the question arises as to how these participants will interact with one another and what strategy they should use to align multiple windows of the immersive environment in a coordinated manner.

\subsection{Summary}
Multiple studies have shown that multiview visualization in two-dimensional screens is useful for tasks involving both sensemaking and exploration. Therefore, it is prudent to investigate how collaborators communicate and interact with multiple windows in collaborative VR environment when the document type varies. The use of automatic layouts can help minimize the view management cost of multiview approaches; however, there are occasions when users may find it more convenient to manipulate views manually; for example while performing a side-by-side comparison. This was proved by related study carried out in the field of immersive analytics that the users positioned information from a primarily self-interested vantage point during the discovery phase \cite{ThereIsNoSpoon, EvaluatingTheBenefits}. This suggests that views are supposed to be directed toward the individual using them. However, it is still not quite clear how pair of collaborators would organize multiscale views and 3D spatial layouts in VR environment.


\section{User Study}



This study explored how pairs interact and collaborate in a virtual environment to perform analytic tasks while managing multiple windows. More specifically, ten groups of two participants remotely interacted with 2D windows containing documents and visual data, working together to organize information and complete analytic tasks. Our research questions include:


\textbf{RQ1}: How should the ideal arrangements of 2D documents/windows be if pair of collaborators interact with them remotely in a VR workspace?

\textbf{RQ2}: In the context of multiview document exploration, how do collaborative behaviors and interaction strategies reform as task complexity changes?

\textbf{RQ3}: For groupthinking and sensemaking schemes, how should a VR workplace be designed? What are the key design aspects that need to be considered for creating interactive multiview supported collaborative workspace in VR?

\subsection{Apparatus and VR App}
The VR collaboration platform was built in Unity (Version 2021.3.3f1) with Photon PUN 2 for real-time multiplayer support. For remote collaboration, we used two VR headsets in separate rooms: an Oculus Rift with a PC (Intel i5-12500, RTX 3060, 16GB RAM) and a standalone Meta Quest 2 (128GB). The dimensions of the tracking area for both locations were 1.5 $\times$ 1.5 meters for each user.

The Unity app is designed to place remote participants in the same virtual room, where they can see each other's avatars and communicate via voice channel. The virtual room features a control panel with five task buttons for five tasks and a ``Save'' button to store window layout positions. For instance, when the ``Naive'' button is pressed, 18 stacked windows appear (\autoref{fig:teaser}b). Participants can manipulate the windows with the controller as needed by grabbing, placing, scaling, rotating, and moving windows. Once they are done manipulating windows, participants can press the ``Save'' to record each window's 3D position, rotation, and scale.

\subsection{Tasks}
We include five types of tasks: (T1) initial layout construction, (T2) search, (T3) comparison, (T4) classification, and (T5) general/final layout construction with three difficulty levels: easy, moderate, and advanced. A similar approach was also used in \cite{AnEmpiricallyDerivedTaxonomy, AnExaminationofGrouping, CollabAR}. Initial(T1) and general(T5) layout construction tasks are consistent across all difficulty levels and serve as the starting and ending tasks for each level. The remaining tasks -- (T2) search, (T3) comparison, and (T4) classification -- vary in complexity for each difficulty level.


\subsubsection{Initial Layout Construction}
At the start of each difficulty level, participants freely arranged 18 square-shaped windows (\autoref{fig:teaser}b) without a specific goal, allowing us to observe window organization in a collaborative immersive space. We call this the ``Naive'' layout. Once they finished, participants explained the reasoning behind their arrangement. The initial layout was then saved and used as the starting point for all subsequent tasks -search, comparison, and classification - at the same difficulty level, maintaining consistency and minimizing bias from initial window positions.


\subsubsection{Search} Search tasks involve examining each window at least once to identify various features. The following sections describe the search tasks for the three different difficulty levels:

\begin{figure*}[!h]
    \centering
    \includegraphics[width=\linewidth]{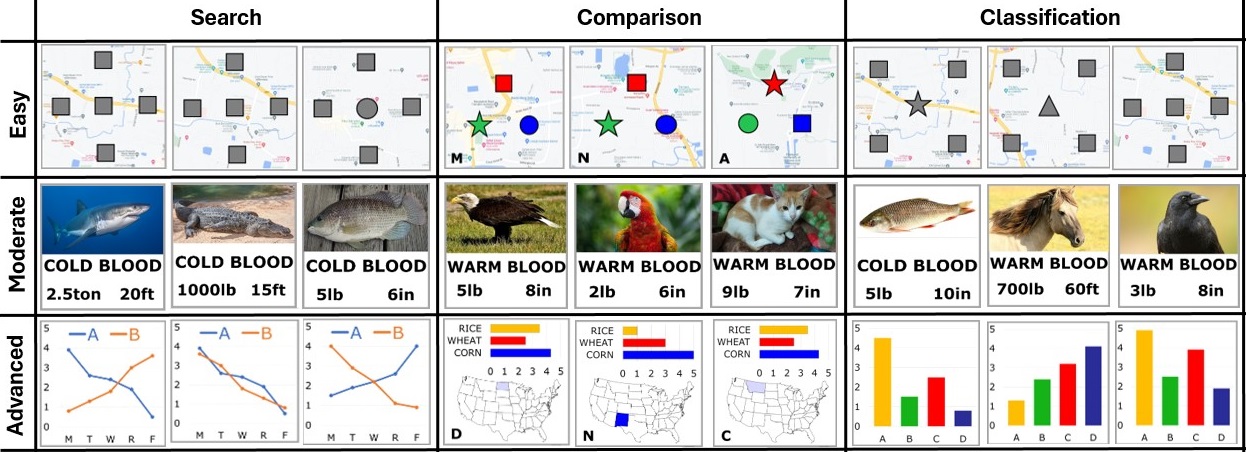}
    \caption{All task variations and their corresponding windows for the study}
    \label{fig:alltasktypes}
\end{figure*}

\textbf{Easy}: Participants were tasked with counting the windows containing a gray circle. At the start, 18 windows were arranged in the virtual space based on the layout created during the initial layout construction phase. Some windows featured one circle and four squares, while others displayed only five squares (\autoref{fig:alltasktypes}, easy search), serving as distractions. For search and other tasks, the placement of these windows within the initial layout was always randomized for each iteration. After completing the search, participants verbally reported their answers to the experimenter.


\textbf{Moderate}: 
The moderate search task required participants to identify specific animals and assess their attributes (e.g., weight) to answer questions involving visual recognition and reading comprehension. Participants were given 18 windows, each displaying various animals (e.g., fish, birds, mammals) along with three attributes: blood type (cold- or warm-blooded), weight, and size (\autoref{fig:alltasktypes}, moderate search). 
To complete the task, they first identified the relevant animals and then assessed their attributes to determine if they met the given criteria - e.g., count windows containing a cold-blooded animal that exceeded a specific weight. This multi-step process added complexity, classifying the task as moderately difficult.


\textbf{Advanced}: The advanced search task required participants to analyze stock price trends across multiple windows to identify specific patterns. Each of the 18 windows displayed two line graphs representing the stock prices of companies A and B over one week (\autoref{fig:alltasktypes}, advanced search). Participants counted the windows where company A's price increased while company B's decreased, requiring them to interpret trends and track occurrences accurately.

\subsubsection{Comparison}
In comparison tasks, participants examined 18 windows to identify similarities and differences between pairs. The following sections describe the comparison tasks for each difficulty level.


\textbf{Easy}: The task required participants to identify pairs of windows that contained matching shapes and colors arranged in identical formations, as used in \cite{mapsAroundMe, ZoomingVersusMultipleWindow}. Each window displayed a square, a star, and a circle arranged in a triangular formation with various colors, also identified with a letter in the bottom left corner. Each task contained one matching pair with identical shapes, colors, and arrangements, while the remaining 16 windows served as distractions. Participants were required to find windows and tell us the letters representing the windows. For example, the first (with letters `M') and second (with letters `N') windows in \autoref{fig:alltasktypes} (easy comparison) share identical shapes, colors, and positioning. 


\textbf{Moderate}: In this task, participants compared animals based on specific attributes, such as weight, after identifying animals of the same species across different windows. We used a new set of 18 windows with animals and their three attributes. The comparison task involved two stages: participants first identified an animal (e.g., birds) and then compared their attributes to find the answer. For example, in \autoref{fig:alltasktypes} (moderate comparison), the first two windows feature bird species, so the weight comparison only occurred between them.


\textbf{Advanced}: The advanced comparison task required participants to identify pairs of U.S. states with identical temperature profiles and crop production data. Each window displayed a state's temperature map and a bar chart of annual rice, wheat, and corn production. Participants compared 18 windows to find matching pairs, as shown in \autoref{fig:alltasktypes} (advanced comparison), where the first and last windows match the temperature and crops production, while the middle one differs.

\subsubsection{Classification}
A classification task involves classifying and making clusters of windows based on pre-defined conditions - as also used in 
 \cite{AnExaminationofGrouping, WhereShouldWePutIt}. We designed easy, moderate, and advanced classification tasks as discussed below.


\textbf{Easy}: In the easy classification task, participants examined 18 windows with geometric shapes and grouped them into three clusters based on the central shape, followed by counting the number of windows in each cluster. Example windows are shown in (\autoref{fig:alltasktypes}, easy classification), where the central shape is a gray circle, triangle, or star. Participants identified windows with the same central shape and grouped them into clusters accordingly. 


\textbf{Moderate}: Participants were given 18 windows displaying animal images and their attributes. They first grouped the animals into three clusters—birds, fish, and mammals—then further categorized them based on specific criteria (e.g., height between 6 and 12 inches) within each cluster (\autoref{fig:alltasktypes}, moderate classification). Finally, they reported the number of windows meeting the criteria in each group.
  


\textbf{Advanced}: In the advanced classification task, participants analyzed 18 windows displaying bar charts of monthly expenditures for four categories: A (employee salary), B (transportation), C (electricity), and D (security) across different companies. They clustered companies with similar spending patterns, as seen in \autoref{fig:alltasktypes} (advanced classification), where the 1st and 3rd windows represent companies with matching expense profiles - e.g., highest spending on A, followed by C, B, and D.

To eliminate learning bias, we used unique images, charts, and information for each window across the three difficulty levels in search, comparison, and classification tasks. This resulted in 162 distinct windows (18 windows $\times$ 3 tasks $\times$ 3 difficulty levels) for the study.

\subsubsection{General/Final Layout}
At the end of each difficulty level, participants were given 18 windows that were constructed after the initial layout construction phase and asked to create a general layout suitable for that level. The goal of this step is to examine how the initial layout evolves into an `optimal' layout that could be used for search, comparison, and classification tasks. We refer to this as a `general' layout. 

\subsection{Participants}
We recruited 20 participants (15 male, 5 female, aged between 21 and 25, M = 23, SD = 1.34) from the local university for the study. None of the participants had prior VR experience. However, 8 participants were familiar with video gaming and controllers. Each participant received compensation for their participation.

\subsection{Procedure}
Upon arriving at the lab, participants provided demographic information and were introduced to the VR headsets and controllers. They received a briefing on the study procedures, tasks, and how to manipulate windows using the controllers (e.g., grabbing, placing, scaling, rotating, and moving). Two researchers guided participants to separate rooms, helped them wear the headsets, and launched the VR application, connecting them to the same virtual room in the VR application, where they appeared side by side. 

\begin{figure}[!h]
    \centering
    \includegraphics[width=.8\linewidth]{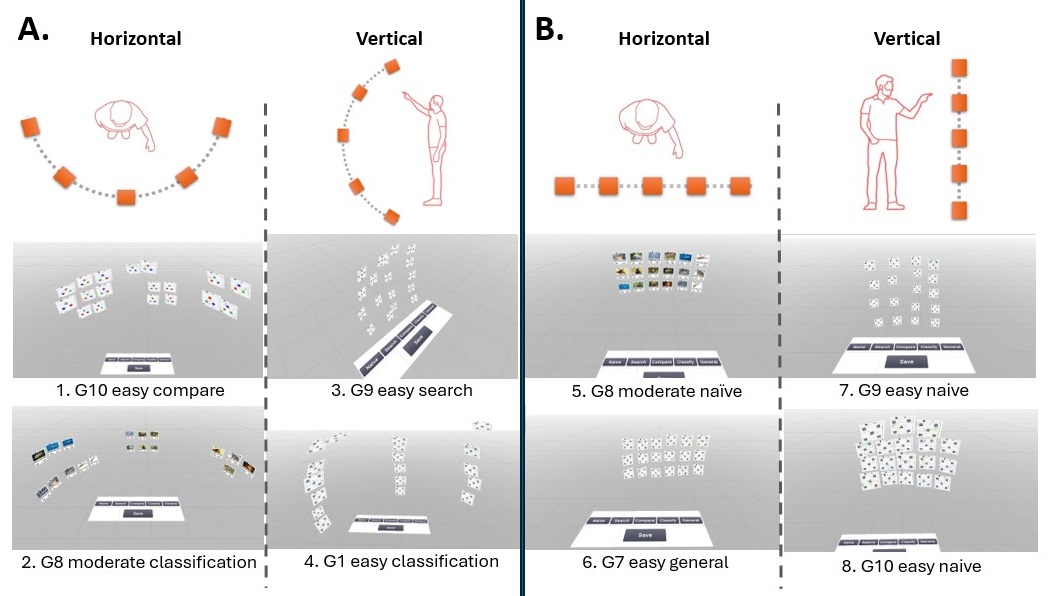}
    \caption{Layout generated by participants: A) Semi-circular layout (vertical and horizontal) and B) Planner Layout (vertical and horizontal)}
    \label{fig:planner&semi}
\end{figure}

Participants were randomly assigned a difficulty level and completed tasks in the following order: initial layout construction, search, comparison, classification, and final layout. In the VR app, they used the ``Naive'', ``Search'', ``Compare'', ``Classify'', and ``General'' buttons to initiate each corresponding task.
In the search, comparison, and classification tasks, participants were instructed to collaboratively organize the windows in a way that would be generalizable and useful for anyone attempting to answer task-related questions. This approach encouraged participants to collaboratively create a structured and organized layout rather than placing windows randomly. At the end of each task, we recorded the position, orientation, and scale of all windows in the 3D environment. 

Each group completed five task types - Initial Layout Construction, Search, Comparison, Classification, and General Layout Construction - at three difficulty levels - Easy, Moderate, and Advanced. With 10 participant groups, we collected a total of 150 window layouts (5 tasks $\times$ 3 difficulty levels $\times$ 10 groups ). The study sessions were video-recorded. After each difficulty level, we collected NASA-TLX data for the search, comparison, and classification tasks and asked participants to explain their reasoning behind the final layouts they designed. Each session lasted 75 minutes.

We used BORIS\cite{Friard2016} to analyze video data. We used open coding of the video material as well as notes that the investigator gathered to analyze the video data. Similar approaches were used in \cite{CollabAR}. We investigated the following collaboration styles:

\begin{itemize}
  \item \textbf{Solo mode}: Solo mode refers to participants focusing entirely on their tasks with minimal to no interaction with remote participants, ensuring full concentration on their own activity.
  
  \item \textbf{Collaborative mode/Active Discussion}: In collaborative mode, participants actively communicate and work together to complete the task. 
  
  \item \textbf{Layout Manipulation}: This refers to the time participants actively interact with windows by selecting, grabbing, rotating, or repositioning them. Layout manipulation can occur in both solo and collaborative modes.
  
  \item \textbf{Conflict}: A conflict occurs when both participants attempt to interact with the same window at the same time during the study. 

\end{itemize}

\section{Results}
\begin{figure}[!h]
    \centering
    \includegraphics[width=.70\linewidth]{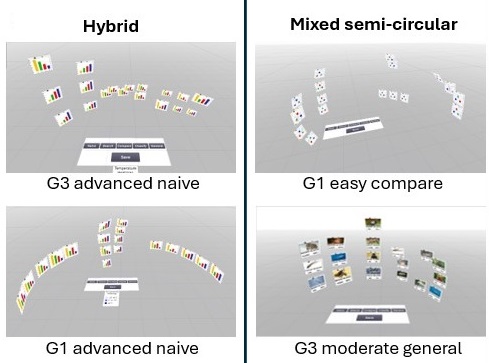}
    \caption{Layout generated by participants: Hybrid and mixed semi-circular}
    \label{fig:hybrid&territory}
\end{figure}

Due to the straightforward nature of the classifications, we used an iterative single-coder method to identify potential layout classes, which was also used in similar studies \cite{effectsOfInteractiveLatency, effectsOfDisplaySize}. Then we refined the classifications through discussions to resolve ambiguities and minimize subjectivity. We labeled interaction strategies and collaboration styles for each group and evaluated task load across task types and difficulty levels using NASA-TLX. A summary of the observed general layouts in the collaborative VR space is provided (\autoref{fig:planner&semi}, \autoref{fig:hybrid&territory}).

\subsection{Layout Geometry}
\label{sec:layoutGeometry}
Out of 150 layouts, we eliminated 7 layouts from 5 groups (G3, G4, G5, G6, and G9) as they did not adhere to any clear pattern. Therefore, the remaining 143 layouts were analyzed.

A thorough inspection of the layouts revealed two distinct geometries: semi-spherical/semi-circular and planar. In the semi-spherical or semi-circular layout, windows are arranged as if placed on a portion of a sphere's surface. Participants positioned windows around themselves in a 360-degree sphere-like layout, creating an immersive workspace that partially enclosed their view. In contrast, the planar layout arranges windows on a flat surface, similar to a traditional screen setup. Participants positioned the windows as they would on a 2D display, aligning them within a single flat plane \cite{mapsAroundMe}. Both the semi-circular and planar layouts appeared in two variations: vertical and horizontal. In the vertical layout, windows were stacked at different heights, requiring participants to look up and down. In contrast, the horizontal layout arranged windows side by side at eye level, allowing for easier left-to-right scanning. As a result, the geometric patterns identified across all 143 layouts can be categorized as follows:

\begin{figure}[!h]
    \centering
    \includegraphics[width=.80\linewidth]{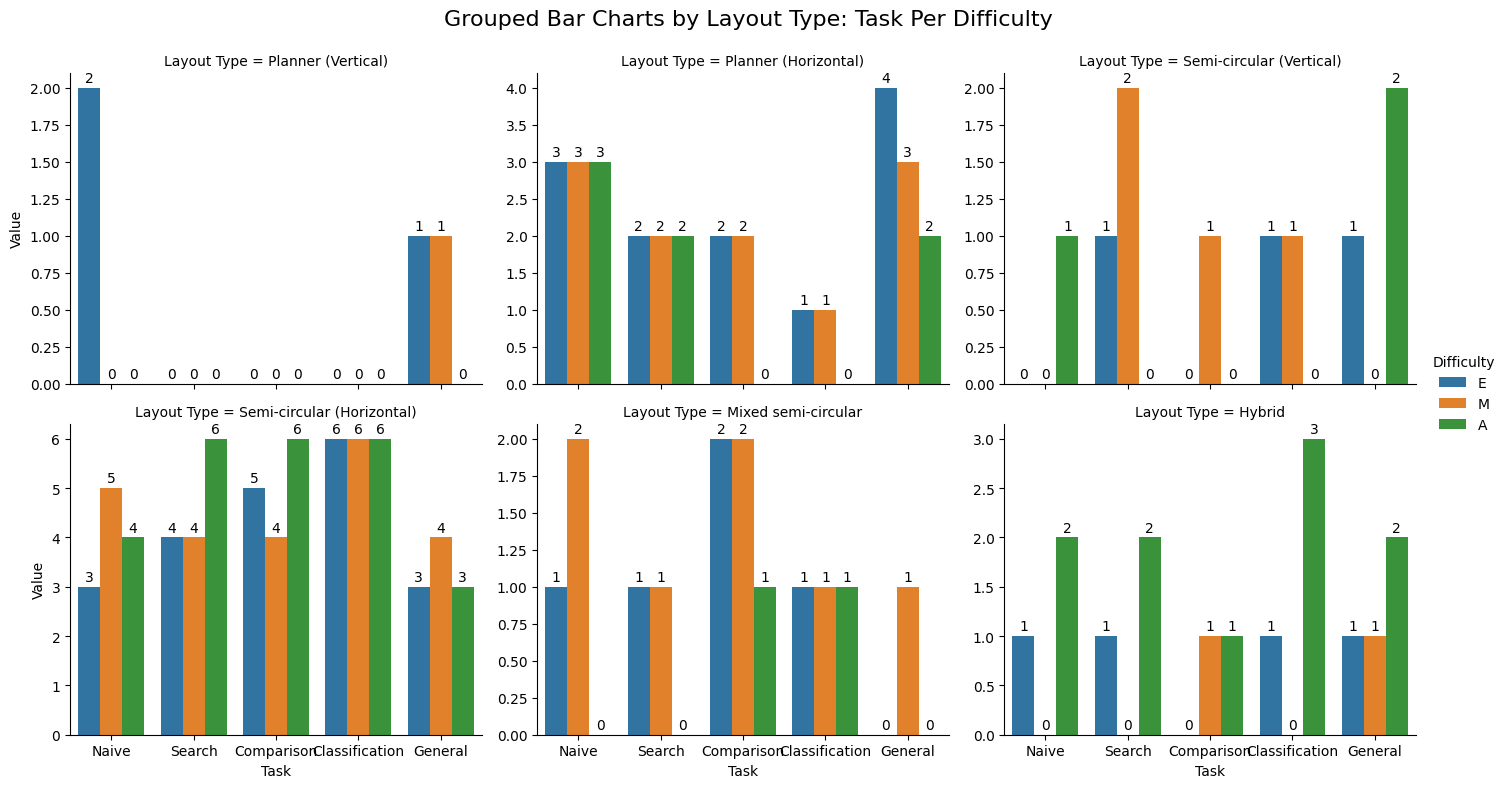}
    \caption{Number of layout patterns generated for each task type and difficulty (E = easy, M = moderate, A = advanced)}
    \label{fig:layoutTable}
\end{figure}





\begin{itemize}
    \item Semi-circular, vertical: 10 layouts (7\%)
    \item Semi-circular, horizontal: 69 layouts (48.25\%)
    \item Mix Semi-circular (vertical + horizontal) : 14 layouts (9.7\%)
    \item Planner, vertical: 4 layouts (2.7\%)
    \item Planner, horizontal: 30 layouts (Mostly on Naive and General) (20.9\%)
    \item Hybrid (vertical planner + horizontal semi-circular) : 16 layouts (11.18\%)
\end{itemize}



The hybrid layout was mostly found in Classify (4) and General Task (4).

In case of difficulty, the usage of planner layout were highest in easy tasks (15) and lowest on Advanced task (7). Semi-circular layouts (vertical, horizontal or mixed) were highest in moderate level tasks(34) and lowest on easy level task (29). Hybrid layout were mostly found on Advanced task (10). But overall the usage of semi-circular layouts were significantly higher then any other layout type in any context. 

\begin{figure*}[t]
    \centering
    {\includegraphics[width=\textwidth]{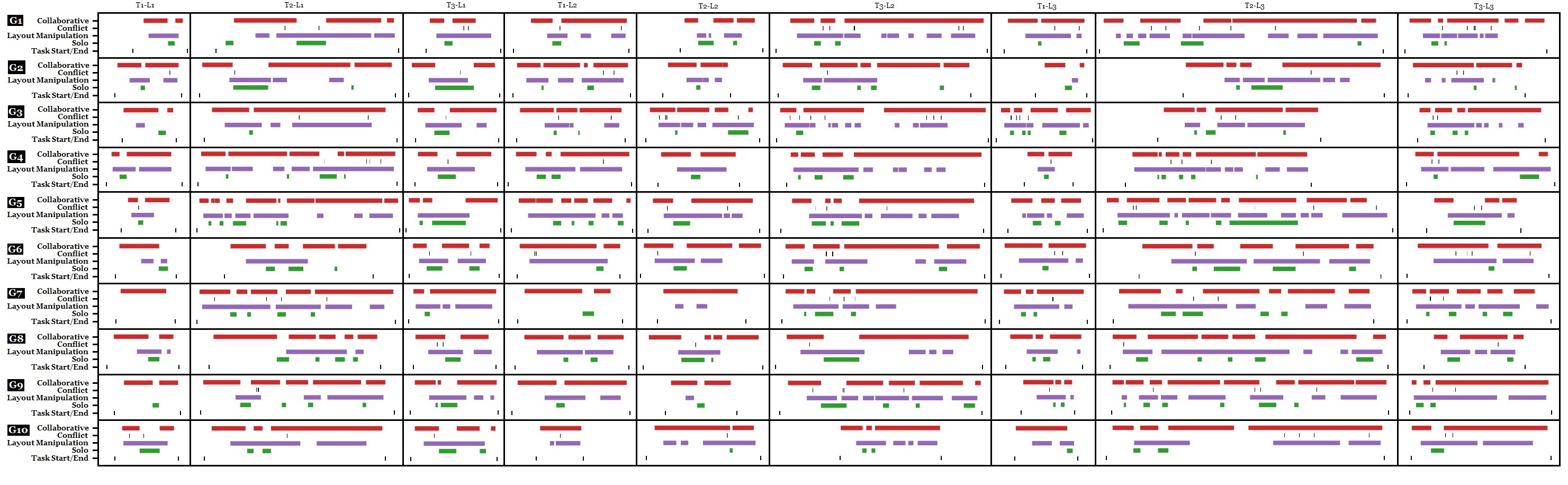}}
    \caption{The timeline shows how each group's video recording was coded in accordance with the identified categories.}
    \label{fig:allgrouptimeline}
\end{figure*}

After analyzing the layouts, we discovered that among 30 sets of tasks (we define a set as the collection of all 5 tasks for a single group in each difficulty level. therefore, 10 groups $\times$ 3 difficulty level = 30 sets), 12 sets began with one layout type in naive task but ended up with a different layout in general task, after going through a complex interaction procedure. This layout changing behaviour was mostly observed in easy level task (6) and least observed in moderate level task (2). Among the 12 sets where layout pattern changed at the end, 5 of them ended up in semi-circular layout. 






\begin{table}[t]
  \centering
  \caption{Mean interaction strategies for search task.}
  \label{tab:interaction-search}
  \resizebox{\columnwidth}{!}{
  \begin{tabular}{lrrrrr}
    \toprule
    \textbf{Interaction strategy} & 
    \textbf{Easy} & 
    \textbf{Moderate} & 
    \textbf{Advanced} & 
    \textbf{F-score} & 
    \textbf{p-score} \\
    \midrule
    Solo mode         & 6.73  & 9.71  & 10.96 & 1.04   & 0.30   \\
    Collaborative mode& 38.18 & 99.88 & 73.80 & 10.06  & 0.0005 \\
    Layout manipulation & 21.45 & 64.47 & 52.22 & 5.76   & 0.008  \\
    Conflict          & 0.40  & 1.00  & 1.10  & 1.417  & 0.259  \\
    \bottomrule
  \end{tabular}}
\end{table}




\begin{table}[t]
  \centering
  \caption{Mean interaction strategies for comparison task.}
  \label{tab:interaction-comparison}
  \resizebox{\columnwidth}{!}{
  \begin{tabular}{lrrrrr}
    \toprule
    \textbf{Interaction strategy} & 
    \textbf{Easy} & 
    \textbf{Moderate} & 
    \textbf{Advanced} & 
    \textbf{F-score} & 
    \textbf{p-score} \\
    \midrule
    Solo mode           & 28.93  & 13.97  & 50.83  & 7.85   & 0.0021  \\
    Collaborative mode  & 177.77 & 83.13  & 269.43 & 14.57  & 0.000051 \\
    Layout manipulation & 126.79 & 56.38  & 217.37 & 16.86  & 0.000018 \\
    Conflict            & 1.60   & 0.90   & 2.90   & 4.95   & 0.0148  \\
    \bottomrule
  \end{tabular}}
\end{table}





\begin{table}[t]
  \centering
  \caption{Mean interaction strategies for classification task.}
  \label{tab:interaction-classification}
  \resizebox{\columnwidth}{!}{
  \begin{tabular}{lrrrrr}
    \toprule
    \textbf{Interaction strategy} & 
    \textbf{Easy} & 
    \textbf{Moderate} & 
    \textbf{Advanced} & 
    \textbf{F-score} & 
    \textbf{p-score} \\
    \midrule
    Solo mode           & 22.34  & 25.30  & 20.12  & 0.44   & 0.649   \\
    Collaborative mode  & 60.31  & 203.34 & 142.97 & 11.91  & 0.000196 \\
    Layout manipulation & 57.35  & 128.11 & 120.14 & 7.05   & 0.0034  \\
    Conflict            & 1.00   & 2.60   & 2.40   & 3.17   & 0.0581  \\
    \bottomrule
  \end{tabular}}
\end{table}


\subsection{Collaboration Strategy}
A thorough breakdown of the group collaboration analysis can be seen in \autoref{fig:allgrouptimeline}. The complete timeline for each group is displayed, with an extensive outline of the various categories, such as conflict, collaboration styles, and task start and end times. The timelines suggest that in general there was a significant number of conflicts across all the participant groups. More precisely, there were 139 Conflicts for 10 groups (avg: 13.9, min: 6, max: 24, std.: 5.63). Interestingly, the number of conflicts for three difficulty-level tasks increased as the task difficulty grew (A total of 30 conflicts for easy, 45 for moderate, and 64 for advanced tasks). An ANOVA test for 3 types of task (search, comparison \& classification) revealed that there was no significant difference between number of conflicts and difficulty levels within the search task (F=1.417, P >0.1). For comparison tasks, we found significant differences across difficulty (F=4.94, P < 0.05). In the case of the classification task, no significant differences were found (F=3.16, P = 0.06, F < F crit).

\begin{figure}[h]
    \centering
    \includegraphics[width=0.7\linewidth]{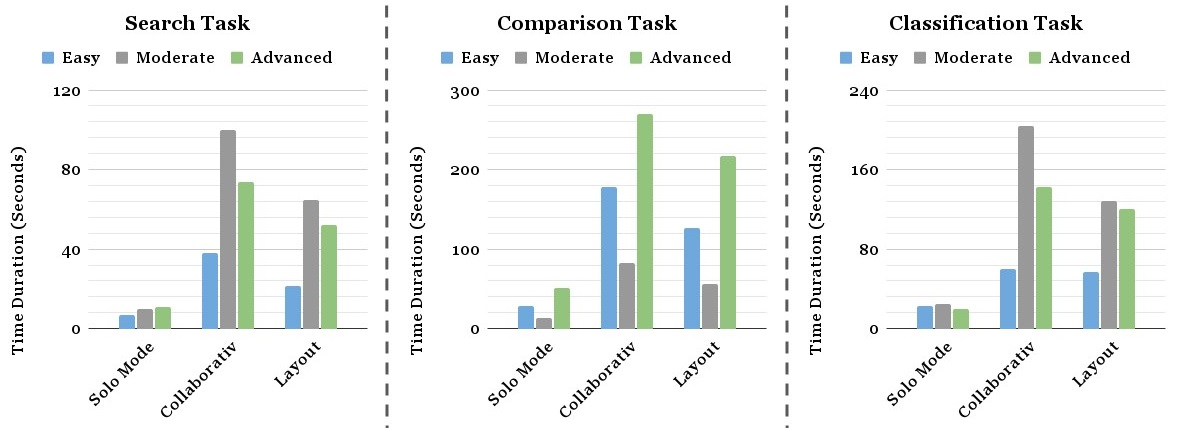}
    \caption{Mean time duration (in seconds) of ten groups for each interaction strategies (solo mode, collaborative mode and layout manipulation) for three task types over three level of difficulty.}
    \label{fig:int_strategies}
\end{figure}

Based on our preliminary findings, we found that the level of complexity of each of our three tasks led to an increase in the amount of information processed and collaboration that took place (\autoref{fig:int_strategies}). After performing the ANOVA test to our data, we found that the variation of collaborative mode durations across different difficulty levels is higher (F=10.06, p < 0.05) for search task. A similar pattern was also found for comparison (F=14.56, p < 0.05) and classification (F=11.90, p < 0.05) task. Post-hoc Tukey HSD Test reveals that for search and classification task, the difference in collaborative mode is significant between easy vs advanced, and between easy vs moderate. For comparison task, difference is significant between all pairs.

\begin{figure}[h]
    \centering
    \includegraphics[width=0.7\linewidth]{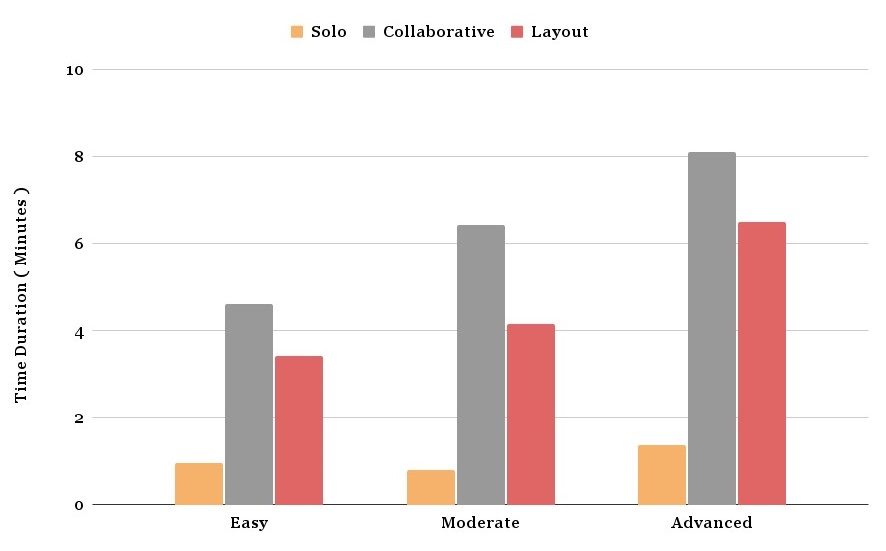}
    \caption{Mean time duration (in minutes) of all tasks for each interaction strategies (solo mode, collaborative mode, layout manipulation) over three level of difficulty.}
    \label{fig:int_strategies_mean}
\end{figure}

There was significant difference in layout manipulation across difficulty levels for search task(F=5.76 and p < 0.05) and classification task (F=7.053 and p < 0.05). The Post-hoc Tukey HSD Test reveals that the difference is significant between easy vs moderate(for search and classification) and between easy vs advanced (for classification). For both search and classification task, the layout manipulation were highest in moderate level difficulty (M=64.47 and 128.11 for search and classification, respectively). For comparison task we also found significant differences across difficulty levels (F=16.858 and p < 0.05). The Post-hoc Tukey HSD Test reveals that the difference is significant between all pairs. But here we observed lowest amount of layout manipulation in moderate difficulty (M=56.3794). Among three types of task, overall we observed large amount of layout manipulation in comparison task.

\subsection{Usability}



\begin{figure}[h]
    \centering
    \includegraphics[width=0.9\linewidth]{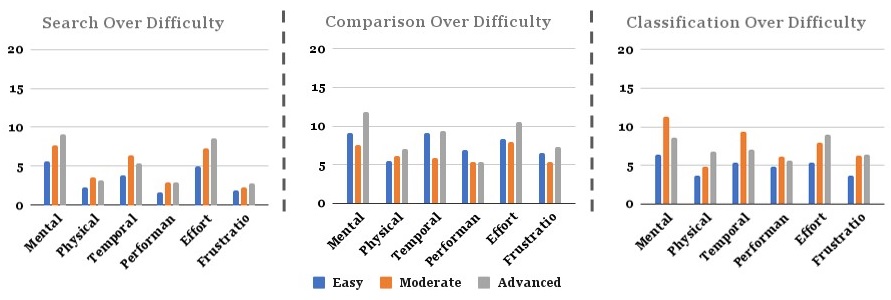}
    \caption{Mean values of NASA-TLX matrices for three tasks over three level of difficulty (Easy, Moderate, Advanced)}
    \label{fig:nasaAllTaskBar}
\end{figure}

Because our tasks require participants to organize their windows in a way that lets them process information and make decisions, the perceived task load over the mental demand and effort matrices of each of the three tasks (search, comparison, and classification) is moderately high (\autoref{fig:nasaAllTaskBar}) at first glance. 

\begin{table}[t]
  \centering
  \caption{NASA-TLX metrics for the search task.}
  \label{tab:nasa-tlx-search}
  \resizebox{\columnwidth}{!}{
  \begin{tabular}{lrrrrr}
    \toprule
    \textbf{NASA-TLX metric} & 
    \textbf{Easy} & 
    \textbf{Moderate} & 
    \textbf{Advanced} & 
    \textbf{F-score} & 
    \textbf{p-score} \\
    \midrule
    Mental       & 5.56 & 7.73 & 9.10 & 10.12 & 0.0005 \\
    Physical     & 2.23 & 3.60 & 3.16 & 2.39  & 0.10   \\
    Temporal     & 3.83 & 6.33 & 5.30 & 6.05  & 0.006  \\
    Performance  & 1.66 & 2.86 & 2.90 & 3.83  & 0.030  \\
    Effort       & 5.00 & 7.33 & 8.60 & 9.93  & 0.0005 \\
    Frustration  & 1.93 & 2.30 & 2.80 & 1.36  & 0.27   \\
    \bottomrule
  \end{tabular}}
\end{table}

Participants reported a noticeable rise in mental workload as the difficulty of each task increased. Statistical analysis revealed significant differences in mental workload across the different difficulty levels for all three types of task (F=10.12 and p < 0.05 for search, F=12.01 and p < 0.05 for comparison, F=10.40 and p < 0.05 for classification). These findings indicate that as the difficulty of the tasks changes, participants experienced a greater mental workload. 

Our findings also indicate that there was no significant variation in physical workload across different difficulty levels in search and comparison task. This suggests that the level of difficulty did not have a substantial effect on the physical demands experienced by participants during these tasks. In the context of the classification task, participants were observed to report a marginal rise in their perceived physical workload as difficulty levels increased. However, it is important to note that these differences did not reach statistical significance.







\begin{table}[t]
  \centering
  \caption{NASA-TLX metrics for the comparison task.}
  \label{tab:nasa-tlx-comparison}
  \resizebox{\columnwidth}{!}{
  \begin{tabular}{lrrrrr}
    \toprule
    \textbf{NASA-TLX metric} & 
    \textbf{Easy} & 
    \textbf{Moderate} & 
    \textbf{Advanced} & 
    \textbf{F-score} & 
    \textbf{p-score} \\
    \midrule
    Mental       & 9.06 & 7.56 & 11.86 & 12.01 & 0.0001 \\
    Physical     & 5.53 & 6.10 & 7.00  & 1.30  & 0.28   \\
    Temporal     & 9.10 & 5.90 & 9.40  & 7.76  & 0.002  \\
    Performance  & 6.93 & 5.36 & 5.30  & 1.58  & 0.20   \\
    Effort       & 8.30 & 7.90 & 10.56 & 9.87  & 0.0006 \\
    Frustration  & 6.53 & 5.33 & 7.33  & 4.30  & 0.020  \\
    \bottomrule
  \end{tabular}}
\end{table}

Upon further analysis of the data, it was also found that there was a significant increase in temporal workload as task difficulty increased across all tasks (F=6.05 and p < 0.05 for search, F=7.76 and p < 0.05 for comparison, F=6.68 and p < 0.05 for classification). This finding suggests that participants experienced a heightened sense of time pressure as tasks became more challenging. These results align with previous research that has shown a positive association between task difficulty and perceived temporal workload. In comparison and classification tasks, participants' performance perceptions remained relatively consistent.

The findings on the F score and the p score for effort in all three tasks indicate that the participants perceived a greater need to invest additional effort when engaging in tasks under conditions that posed significant challenges. In addition, participants did not indicate a statistically significant increase in their levels of frustration when engaging in the search task. However, participants exhibited increased levels of frustration when tasked with classifying and comparing data under more difficult conditions.



\begin{table}[t]
  \centering
  \caption{NASA-TLX metrics for the classification task.}
  \label{tab:nasa-tlx-classification}
  \resizebox{\columnwidth}{!}{
  \begin{tabular}{lrrrrr}
    \toprule
    \textbf{NASA-TLX metric} & 
    \textbf{Easy} & 
    \textbf{Moderate} & 
    \textbf{Advanced} & 
    \textbf{F-score} & 
    \textbf{p-score} \\
    \midrule
    Mental       & 6.40 & 11.33 & 8.56  & 10.40 & 0.0004 \\
    Physical     & 3.73 & 4.80  & 6.80  & 3.97  & 0.030  \\
    Temporal     & 5.36 & 9.33  & 7.03  & 6.68  & 0.004  \\
    Performance  & 4.90 & 6.13  & 5.56  & 0.40  & 0.60   \\
    Effort       & 5.40 & 7.90  & 9.03  & 8.94  & 0.001  \\
    Frustration  & 3.73 & 6.26  & 6.43  & 4.57  & 0.010  \\
    \bottomrule
  \end{tabular}}
\end{table}


\section{Discussion}

\subsection{Layout Geometry}
We found that groups tend to arrange their views in a semi-circular (vertical, horizontal or mixed) arrangement, which is an interesting finding considering the prominence of flat whiteboards and enormous display panels in the actual world. In addition, while some planar and hybrid arrangements were made, there was a larger inclination toward semi-circular layouts in general, and no fully spherical arrangement were found. This is likely because when in collaborative environment, semi-circular layouts enable users to examine the overall pattern with better ease while eliminating the need to swivel one's body, and considering the contents on the edge of a layout, semi-circular layouts are easier to visualize then planar as they are closer to the user. 
Interestingly, we found a new type of layout pattern, hybrid, that haven't been discovered yet. This layout pattern specifically can have a good potential when working in a group of two, because of it's way of arranging the windows (\autoref{fig:hybrid&territory}, A) towards the two collaborators. A central planner formation with semi-circular arrangement on both side might bring a great potential in efficiency, information sharing and brainstorming.

There was a gradual increase in the percentage of semi-circular layouts throughout the task, specially when participants had to focus more on the window to complete the task (search, comparison, classify). As the difficulty level increased, we noticed reduction in usage of a planner layout and increase of semi-circular and hybrid layout. This observation suggest that as the task complexity increases, the windows around the groups need more focus and involvement from them, which leads them to switch into hybrid or semi-circular layouts. The highest amount of semi-circular layouts were found in moderate level, suggesting that when working with image+text documents in a collaborative setup, the semi-circular layout can bring the highest level of information visualization and processing.

We initially presented the windows in a stacked formation, provided simple and easy-to-use controls, including depth and pivot manipulations to freely place the windows, and conducted the research in an environment that was devoid of any external reference points. These are some of the many steps that we took to reduce bias in the design of our study and minimize the influence it had on the layouts that participants produced. Nonetheless, our interface design decisions may have contributed to the prevalence of semi-circular layouts. In instance, the handheld controller served as the pivot point for window layout, so participant-facing windows were always the default orientation. In order to create a flat arrangement, users would have had to walk around or manually adjust the angle of the windows using the controller, both of which cost time and effort. It is possible that participants considered the window rotation option too difficult to operate; nonetheless, this feature required the same directional pad as the regularly used depth-displacement tool.


\subsection{Collaboration Strategy}
For comparison task, we find a significant difference in number of conflicts across difficulty levels. We conclude that when doing comparison-like tasks, each user needs to move windows close to another to make the comparison and decision. Hence as task complexity increases, it requires large movements of the windows to compare every pair side by side. At the same time the task nature makes it difficult to split the task among the collaborators, unlike search or classification task. Thus users tend to move the windows randomly without any proper separation, resulting large conflict.
According to our result, the interaction methods that participants use to complete our tasks have a considerable effect according to the level of difficulty. The higher the level of complexity, the more time the participants spend inside their groups collaborating and manipulating the layout. Solo mode has a less impact on participants to perform tasks of different difficulty level. The reason for this might be that in order for the participants to successfully complete each task and answer all of our questions, they need to work together and interact with windows for brainstorming purposes more than they do when working alone.

For search and classification task, the amount of layout manipulation was highest in moderate level. It might suggest that graphs and shapes related documents require less window manipulation to perform any search or classification related task. In contrast, for comparison related works, image+text documents might require less window manipulation to get the work done in collaborative VR environment. Similar conclusion is also found from analyzing task time.

\subsection{Usability}
Our study findings indicate that participants consistently reported heightened mental workload, temporal demands, and effort as the level of task difficulty increased across all three task types, namely search, comparison, and classification ( \autoref{tab:nasa-tlx-search}, \autoref{tab:nasa-tlx-comparison} and \autoref{tab:nasa-tlx-classification}). That means collaborative interaction, information gain from texts, images and graphs, finding answer to specific questions after analyzing gained information become harder as difficulty level grows from easy to advanced. This pattern  satisfies our division of each task into three level of difficulty. In order to effectively categorise windows and derive meaningful conclusions, participants are required to engage in a meticulous examination of each window. By applying their own cognitive frameworks and subjective perspectives, participants can discern patterns and similarities among the windows, facilitating the process of grouping them. This analytical endeavour serves as a means to ultimately uncover the desired answers or insights.




Interestingly, for comparison and classification task, we observe that the mental and temporal demand of moderate level difficulty is different then what we expect. For comparison task the scores are lower then easy level, and for classification task scores are high then advanced level. This can lead to a conclusion that user feels less stressed when doing comparison related tasks with image+text documents, and graphical visualization of data requires less mental and temporal demands in classification related task. Similar conclusion is also found by analyzing collaboration strategies.


\subsection{User Experience and Observation}


Initially participants observed their assigned task in their own-created naive layout. Then, they meticulously inspected each window individually to identify general patterns. We characterize this style of interaction as solo mode in which individuals independently modify windows and locate information without communicating with their respective teammates. Participants then discussed and attempted to divide their tasks into smaller portions in order to obtain data from windows (collaborative mode). Then they worked independently (solo mode) for a period of time. For example, at first participants together explored how to sort and find information from windows based on text and graphics. Having deciding on this course of action, they performed independently without interacting. Then, after finding some information, they shared it with their partner and continued to search for an answer. If they were unable to locate the correct response to our question, they repeated the previous steps using solo mode, collaborative mode, and layout manipulation. Before completing a layout, group members confirmed each other's opinions. 

Based on our findings, the collaborative mode is the most prominent of the three interaction modes. Participants spent 5 to 10 times longer in collaborative mode than solo mode for the search task (\autoref{fig:int_strategies_mean}). The same pattern of higher collaborative mode can also be observed for comparison and classification tasks, ranging 3-8 times then solo mode. After collaborative mode, layout manipulation is the second most prevalent interaction strategy, indicating that participants spent the most time on layout modification. 





\subsection{Design Guidelines}
To be able to move the field ahead, it is vital to have a solid understanding of how virtual reality interfaces affect remote collaboration. New design ideas and studies are required in order to pave the way for new frameworks for collaborative virtual reality, which are necessary in order to consolidate a variety of VR applications. Our research focuses on various aspects of VR collaboration. Through the measurement of collaboration parameters such as layout geometry, task loads, collaboration styles and conflicts, we deliver insight and perspective of using multiview UI windows in VR collaboration. The following design principles are ones that we think should be used to improve the experience of remote VR collaboration. They are directly based on the empirical facts that we gathered and are geared toward the development of systems and applications for collaborating on projects.

\textbf{Use semi-circular Layout for Collaboration:}
We find large amount of semi-circular arrangements in our study. Groups that came up with this layout pattern mentioned in the interviews that they intended to come up with an arrangement that would let them examine all of the windows at once, also being close to them. Because of the relatively high number of the windows, arranging them all on a plane might result in a skewed perspective of the windows that are located along the layout's perimeter. The windows can be viewed from the optimal angle by all thanks to a semi-circular arrangement, which also reduces the amount of movement that users need to do in order to navigate the environment by rotating their bodies or moving about. We also suggest to explore the hybrid layouts because it can offer unique functionalities and separation of objectives.

\textbf{Awareness Cues:}
We notice a significant number of conflicts in our study. To reduce the conflicts and further improve remote VR sensemaking and brainstorming experience, more visual helping functions, such as annotating and creating visual linkages, should be added. Visual input may be offered to users on where other users are looking and who is interacting with which window in view. Collaborative VR interfaces have the potential to provide awareness cues for the content that is being used by each individual, much like screen-based groupware does. One illustration of this is the `turn-based' colored frame visualization that can be found in CollabAR\cite{CollabAR} ; however, the interface may also incorporate more complex visual signals.

\textbf{Automatic Layout Mechanism and Window positioning:}
A range of preconfigured layouts may provide early assistance for many types of collaborative tasks. We notice significant amount of window movement and manipulation, specially in comparison task. Hence, a system where a grabbed window will automatically stick to the intended position on the layout after releasing should reduce the time, effort and task load drastically. 
Consequently, it is essential that immersive multiview interface systems offer manual window layout modifications. These manual features should provide choices for moving both individual windows and groups, such as moving all windows together or doing a rapid examination without modifying the layout. Maintaining the user's mental representation of the system also requires a seamless transition from one layout to another and the ability to return to specified or bookmarked layouts.

\textbf{Smooth Scaling Mechanism for Individual Windows and Fonts:}
It was observed in out study that participants tends to bring the windows closer to visualize better, instead of increasing the size. This could lead to a conclusion that, the size changing mechanism with buttons is not a best implementation, and should be avoided. We recommend to implement the scaling functionality with some smooth gesture that is easy to apply for users. Also, for document and graph visualization, the font size plays a vital role; too small font size forces the participants to keep the windows closer. We suggest to add font size changing functionality to the layout interfaces rather then keeping it fixed.
\section{Limitations and Future Work}
Despite the fact that our findings provide insightful suggestions on how diverse layouts are formed in a collaborative VR environment for sense-making activities, our work has some limitations. 

Our study was small in both size and scope, with only ten groups, and the demographics of those participants were not very diverse. Even though we made an effort to establish an atmosphere that reduces bias in layout creation (an infinite free space), the study was conducted in a controlled setting without a particular emphasis on any one application area. Also, none of the participants had any previous experience with VR. If we could conducted the study with participants having diverse demographics, then we would probably uncover some more interesting insights. The study was conducted with 2 participants in each group. It would be interesting to see how the layout changes as the participants in each group increases gradually.


In our study, the VR room for collaboration were totally in virtual free space, with no walls or boundaries in it, so that all the potential bias can be removes. An interesting area to explore would be to observe and compare the interaction strategy and design pattern of the layouts contracted in a remote room-scale VR vs infinite scale VR, similar to the previous work in AR\cite{WhereShouldWePutIt}.

While designing task for our study, we noticed that the font size played a vital role for visualizing and analysing the graphs and documentations. When we designed our tasks for the first time, participants had to bring the windows much closer, or scale them significant amount in order to analyse them, as our font choice was too small at the first time. We re-designed all the task again to calibrate the font size so that no scaling bias occurs for too small fonts. A future work might include a deep investigation of the effect of font size on document placement and layout formation strategies for graph and document visualization in VR environment.

\section{Conclusion}
Our experiment on remote VR collaboration explores multiview layout management and users' interaction techniques. Our research's objective is to comprehend how users set up their surroundings in order to manage immersive multiview window layouts when in remote collaboration. The findings of our research make it possible for us to recognize unique patterns of layout geometry (Semi-circular, planar, hybrid), collaboration and interaction strategies and identify subjective task load assessment related to task difficulty.

Our primary findings are that groups prefer and organize multiview windows in a semi-circular shape around them, and that they frequently reorganize the windows and face conflicts while they work. Results from recorded video analysis to determine interaction techniques (solo mode, collaborative mode and layout modification) indicate that participants spend the most time interacting collaboratively, which implies they spent more time discussing and brainstorming with their respective teammates. Our Nasa-TLX data suggests that relatively difficult tasks involve greater mental demand and effort. The overall task load index is moderately high, supporting our classification of each work into three categories of complexity. Analyzing the task load and task completion time we also find that for comparison related task, documents with text+image performs better and for classification related task, graphical visualization requires less mental and temporal demand in multi-window system.

Thus, we can conclude that a greater degree of collaborative interaction facilitates our task design for determining layout management and interaction techniques in a collaborative remote environment. And on the basis of these findings, we propose some additional design recommendations that can help steer the creation of VR sensemaking and group discussion system.





\bibliographystyle{ieeetr}
\bibliography{reference.bib}

\end{document}